\input{aipcheck}

\documentclass[
    ,final            % use final for the camera ready runs
  ]
  {aipproc}

\layoutstyle{6x9}
\usepackage{amsmath,amssymb}

\begin{document}

\title{Electroweak tests of the Standard Model}

\classification{12.15.-y,14.60.-z,14.70.-e,14.80.Bn.}
\keywords      {Electroweak precision data, global fit, Higgs boson mass.}

\author{Jens Erler}{
  address={Departamento de F\'isica Te\'orica, Instituto de F\'isica, 
  Universidad Nacional Aut\'onoma de M\'exico, M\'exico D.F. 04510, M\'exico}
}

\begin{abstract}
Electroweak precision tests of the Standard Model of the fundamental interactions are reviewed
ranging from the lowest to the highest energy experiments. 
Results from global fits are presented with particular emphasis on the extraction of fundamental
parameters such as the Fermi constant, the strong coupling constant, the electroweak mixing angle, 
and the mass of the Higgs boson. 
Constraints on physics beyond the Standard Model are also discussed.
\end{abstract}

\maketitle

%%%%%%%%%%%%%%%%%%%%%%%%%%%%%%%%%%%%%%%%%%%%
%% MAINMATTER
%%%%%%%%%%%%%%%%%%%%%%%%%%%%%%%%%%%%%%%%%%%%

\section{Introduction}

The Standard Model (SM) of the electroweak (EW) interactions has been developed
mostly in the 1960s, where the gauge group $SU(2)_L \times U(1)_Y$ was 
suggested~\cite{Glashow:1961tr}, the Higgs mechanism for spontaneously broken
gauge theories developed, and the model for leptons constructed 
explicitly~\cite{Weinberg:1967tq}.  Subsequently, key predictions of the SM were
observed in the 1970s, including neutral currents and parity non-conservation 
in atoms and in deep-inelastic electron scattering~\cite{Prescott:1978tm}.
The basic structure of the SM was established in the 1980s after
mutually consistent values of the weak mixing angle, $\sin^2 \theta_W$,
were determined from many different processes.
The 1990s saw the highly successful $Z$-factories, LEP and SLC, and 
the confirmation of the SM at the loop level.  It thus became clear that 
any new physics beyond the SM could at most be a perturbation.
The \hbox{previous} decade added precision measurements in the neutrino and quark sectors 
(including a 0.5\% measurement of the top quark mass~\cite{Aaltonen:2012ra}), 
as well as ultra-high precision determinations of the $W$-boson mass, $M_W$ 
(to $2\times 10^{-4}$)~\cite{MW}, the anomalous magnetic moment of the muon~\cite{gminus2},
and the Fermi constant, $G_F$~\cite{mulan}.
These results suggest that the new physics must be separated by at least a little hierarchy 
from the EW scale unless one considers the possibility that a conspiracy is at work.
The current decade will elucidate the EW symmetry breaking sector at the LHC
and witness a new generation of experiments at the intensity frontier with sensitivities to
the multi-TeV scale and beyond. 

%The next section reviews some recent developments in slightly more detail.
%Interpretations of these and other results for the mass of the Higgs
%boson, $M_H$, and new physics, are discussed, respectively, in the two sections thereafter.

\section{Recent Developments}

The MuLan Collaboration at the PSI in Switzerland~\cite{mulan} has measured the $\mu$-lifetime
to parts-per-million precision,
$\tau_\mu = 2.1969803(2.2) \times 10^{-6} \mbox{ s}$, which translates into a determination of
\begin{equation}
G_F = 1.1663787(6) \times 10^{-5} \mbox{ GeV}^{-2}.
\end{equation}
The Higgs vacuum expectation value is given by
$\langle 0|H|0\rangle= (\sqrt{2} G_F)^{-1/2} = 246.22$~GeV.
This result is so precise that even the error in the definition of the atomic mass unit (u) 
can shift $G_F$
(MuLan quotes $G_F = 1.1663788(7) \times 10^{-5} \mbox{ GeV}^{-2}$).
Moreover, the effect of the finite $M_W$ in the W-propagator is no longer negligible.
One may either correct for it, {\em i.e.,} absorb it in $\Delta q$ defined through 
$\tau_\mu^{-1} \propto G_F^2 m_\mu^5 (1+ \Delta q)$, 
or else~\cite{vanRitbergen:1999fi} absorb it in $\Delta r$~\cite{Sirlin:1980nh} defined 
in terms of the accurately known 
fine structure constant, $\alpha$, and $Z$-boson mass, $M_Z$, 
\begin{equation}
  \sqrt{2} G_F M_W^2 \left( 1 - {M_W^2\over M_Z^2} \right) \equiv {\pi\alpha\over 1- \Delta r}\ .
\end{equation}
The latter convention is motivated by an effective Fermi theory point of view, 
and used by MuLan and since this year also by the PDG~\cite{PDG2012}.

What $\tau_\mu$ is to $G_F$ is the $\tau$-lifetime to the strong coupling constant, $\alpha_s$.
At least one low-energy $\alpha_s$-value is needed to promote the $Z$-width and related $Z$-pole
observables from a quantitative measurement in QCD to an EW SM test 
(or to constrain physics beyond the SM). 
Perturbative QCD has recently been extended to 4-loop order~\cite{Baikov:2008jh}, 
but there is a controversy whether the perturbative series should be 
truncated (FOPT)~\cite{Beneke:2008ad} or whether higher order terms from the running 
strong coupling in the complex plane should be re-summed (CIPT)~\cite{Le Diberder:1992te}.
There are also non-perturbative contributions parametrized by condensate terms 
which can be constrained by experimentally determined spectral functions.
There are two different approaches~\cite{Davier:2008sk,Boito:2012cr}
which at present give very similar numerical results.
Using FOPT and the condensates from Ref.~\cite{Boito:2012cr} we find,
\begin{equation}
\alpha_s [\tau] = 0.1193 \pm 0.0021, \hspace{72pt}
\alpha_s [Z] = 0.1197 \pm 0.0028,
\end{equation}
where the latter determination from  the $Z$-pole is the only extraction of $\alpha_s$ with a very 
small theory uncertainty.  
The two values can be seen to agree perfectly.

The most precise derived and purely EW precision observable is no longer the $Z$-pole combination of 
$\sin^2\theta_W$, but rather $M_W = 80.387 \pm 0.016$ GeV from the CDF and D\O\ Collaborations
at the Tevatron~\cite{MW} which is dominated by a $\pm19$~MeV determination by CDF using only
2.2~fb$^{-1}$ of their data.
Together with the LEP~2 combination~\cite{Alcaraz:2006mx}, $M_W = 80.376 \pm 0.033$~GeV, one 
obtains for the on-shell definition of $\sin^2\theta_W$,
\begin{equation}
\sin^2\theta_W^{\rm on-shell} \equiv 1 - {M_W^2\over M_Z^2} = 0.22290 \pm 0.00028,
\end{equation}
from which and can extract $M_H = 96^{+29}_{-25}$~GeV.  As for the updated global EW fit, we find,
\begin{equation}
M_H = 102^{+24}_{-20} \mbox{ GeV}.
\end{equation}
The prospects for the full 10~fb$^{-1}$ dataset are a $\pm 13$~MeV $M_W$ determination
from CDF alone, even when no reduction of the parton distribution function ($\pm 10$~MeV) 
and QED ($\pm 4$~MeV) uncertainties is assumed.
In the most optimistic scenario, CDF could shrink the error to $\pm 10$~MeV,
which is to be compared with the $\pm 6$~MeV accuracy expected from a threshold scan
at a future International Linear Collider.

The anomalous magnetic moment of the muon was measured to extreme precision,
\begin{equation}
\label{amu}
a_\mu \equiv {g_\mu - 2\over 2} = (1165920.80 \pm 0.63) \times10^{-9},
\end{equation}
by the BNL--E821 Collaboration~\cite{gminus2}.
The prediction,
$a_\mu = (1165918.41 \pm 0.48) \times 10^{-9}$,
from the SM includes $e^+ e^-$ as well as $\tau$-decay data in the dispersion integral
needed to constrain the two- and three-loop vacuum polarization contributions and
differs by $3.0~\sigma$.
The data based on $\tau$-decays requires an isospin rotation and a corresponding
correction to account for isospin violating effects and suggest a smaller ($2.4~\sigma$)
discrepancy, while the $e^+ e^-$-based data sets (from annihilation and radiative returns)
by themselves would imply a $3.6~\sigma$ conflict.  
Indeed, there is a $2.3~\sigma$ discrepancy between the experimental branching ratio,
$B(\tau^- \to \nu \pi^0 \pi^-)$, and its SM prediction using the $e^+ e^-$ data~\cite{Davier:2010nc}.
In view of this, it is tempting to ignore the $\tau$-decay data and blame the difference
to the $e^+ e^-$ data on unaccounted for isospin violating effects. 
However, there is also a $1.9~\sigma$ experimental conflict between KLOE and BaBar 
(both using the radiative return method~\cite{Arbuzov:1998te})
the latter not being inconsistent with the $\tau$-data.
As for the question whether the deviation in $a_\mu$ may arise from physics beyond the SM
(especially supersymmetry), my personal take is that  I am less concerned about these hadronic 
issues than the absence of convincing new physics hints at the Tevatron or the LHC.
In any case there is an important new proposal at Fermilab to improve on the precision
in Eq.~(\ref{amu}) by a factor of four~\cite{gminus2}.

\begin{figure}
  \includegraphics[height=.46\textheight]{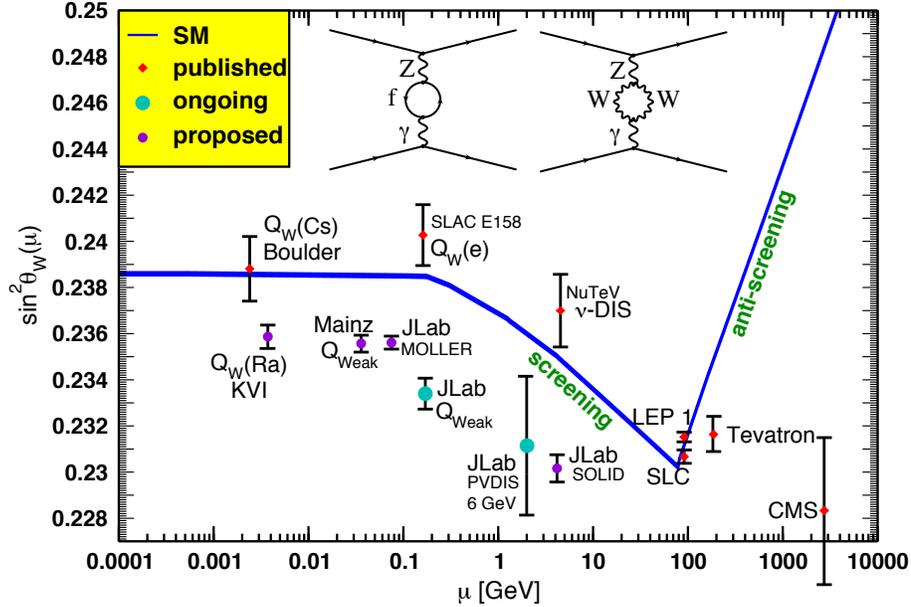}
  \caption{Current and future measurements of the running weak mixing angle. 
  The uncertainty in the prediction is small except possibly in the hadronic transition region roughly
  between 0.1 and 2~GeV~\cite{Erler:2004in}.  
  The relevant $Q^2$ of the Tevatron and CMS values make them effectively additional $Z$-pole
  measurements, but for clarity they have been shifted horizontally to the right.
\label{sin2theta}}
\end{figure}

\subsection{Parity-violating electron scattering}
High precision measurements in the EW sector are also possible at the intensity frontier,
when QED and QCD effects are filtered out by using parity-violating observables.

The JLab Qweak detector~\cite{Qweak} at the 6~GeV CEBAF was dedicated to a measurement 
of the weak charge of the proton, $Q_W^p \propto 1-4 \sin^2\theta_W$, to 4\% precision 
in elastic polarized $e^- p$ scattering at $Q^2 = 0.026$~GeV$^2$.
Data taking is complete and the analysis is in progress.
$Q_W^p$ is similar to the weak charges of heavy nuclei measured in atomic parity violation (APV) 
but at a different kinematics. 
This circumstance results in a re-enhancement of the $\gamma-Z$ box contribution~\cite{gammaZ}
introducing an extra theory uncertainty. 

%The $\gamma-Z$ box is less of an issue at lower $Q^2$ which is one of the reasons why a similar 
%experiment is also planned at a future facility (MESA) in Mainz at $Q^2 = 0.0022$~GeV$^2$. 
%The projected uncertainties for $Q_W^p$ and the extracted $\sin^2\theta_W$ are 2.1\% and 
%$\pm 0.00037$, respectively.

\begin{figure}
  \includegraphics[height=.46\textheight]{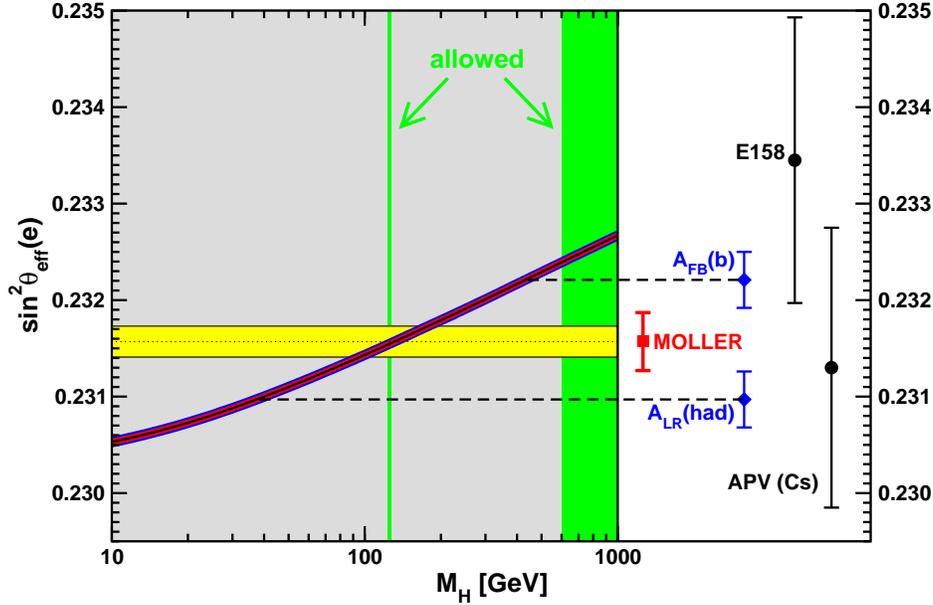}
  \caption{Implication of $\sin^2\theta_W$ measurements for $M_H$. 
  Shown are the most precise determinations from LEP~1 and the SLC, and the extractions from APV
  and from current (E158) and future (MOLLER) polarized M\o ller scattering.
  Also indicated are the non-excluded intervals from direct Higgs searches.
\label{KK}}
\end{figure}

MOLLER~\cite{MOLLER} is an ultra-high precision measurement of $\sin^2\theta_W$ 
in polarized M\o ller scattering at the 12~GeV upgraded CEBAF~\cite{12GeV}.
It aims at a factor of 5 improvement over a similar experiment at SLAC 
by the E158 Collaboration~\cite{Anthony:2005pm}, and would be one of the worlds most precise 
determinations of $\sin^2\theta_W$ and the most accurate at low energies. 

PVDIS was a deep-inelastic polarized $e^-$ scattering experiment using the 6~GeV CEBAF
and is currently in the analysis phase~\cite{PVDIS}. Together with SOLID (at 12~GeV) 
an array of kinematics points will be measured to test strong, EW, and new physics.

Fig.~\ref{sin2theta} summarizes these and other current and future (projected) determinations
of $\sin^2\theta_W$ as a function of energy scale $\mu$.

\begin{figure}
  \includegraphics[height=.46\textheight]{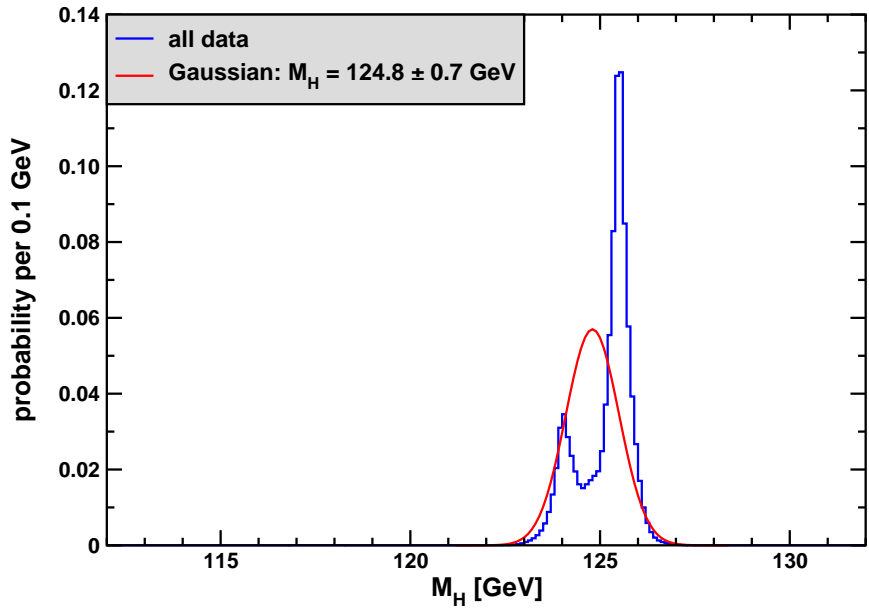}
  \caption{The histogram shows the normalized probability distribution of $M_H$.   
  The bell shaped curve is a reference Gaussian density defined to contain the same probability 
  as the histogram over the region of bins which are higher than the tail bins. 
  The significance of this region corresponds to $3.4~\sigma$.
 \label{mh}}
\end{figure}

\section{SM Interpretation: $M_H$}
The various $\sin^2\theta_W$ measurements discussed above can be used to constrain $M_H$ 
and compare it with the results obtained at the LHC.  
It is important to recall that the most precise determinations at LEP~1 
(from the forward-backward cross-section asymmetry of $Z$-bosons 
decaying into $b\bar b$ pairs, $A_{FB}(b)$) and at the SLC 
(from the polarization asymmetry for hadronic final states, $A_{LR}({\rm had})$), 
both of which being mostly sensitive to the initial state (electron) coupling, 
are discrepant by three standard deviations. 
Their average, on the other hand, corresponds to values of $M_H$ that are in perfect agreement
with the Higgs boson candidates seen by the ATLAS~\cite{:2012an}
and CMS~\cite{Chatrchyan:2012tx} Collaborations at the LHC. 
This is illustrated in Fig.~\ref{KK} together with the low-energy determinations from 
E158~\cite{Anthony:2005pm} and APV which is dominated by the experiment in 
Cs~\cite{Wood:1997zq} and makes use of the atomic theory calculation\footnote{After the conference
had adjourned there appeared an update of the atomic structure calculation~\cite{Dzuba:2012kx} 
finding significant corrections to formally subleading terms.
Taking this into account moves the extracted Cs weak charge $1.5~\sigma$ below the SM 
prediction, which then favors lower values of $M_H$.} 
of Ref.~\cite{Porsev:2010de}.

Estimating the significance of the LHC data~\cite{:2012an,Chatrchyan:2012tx} by themselves 
requires a "look elsewhere effect correction" which is, however, poorly defined.  
It can be avoided when they are combined with the Higgs search results 
from LEP~2~\cite{Barate:2003sz} and the Tevatron~\cite{TEVNPH:2012ab} as
well as with the EW precision data~\cite{Erler:2012uu}, 
the latter providing a normalizable probability distribution (shown in Fig.~\ref{mh}).
This requires the validity of the SM which used to be a very strong assumption in the past.
But with the absence of clear new physics signals at the energy frontier 
this can now be seen as a reasonable approximation.

\begin{figure}
  \includegraphics[height=.46\textheight]{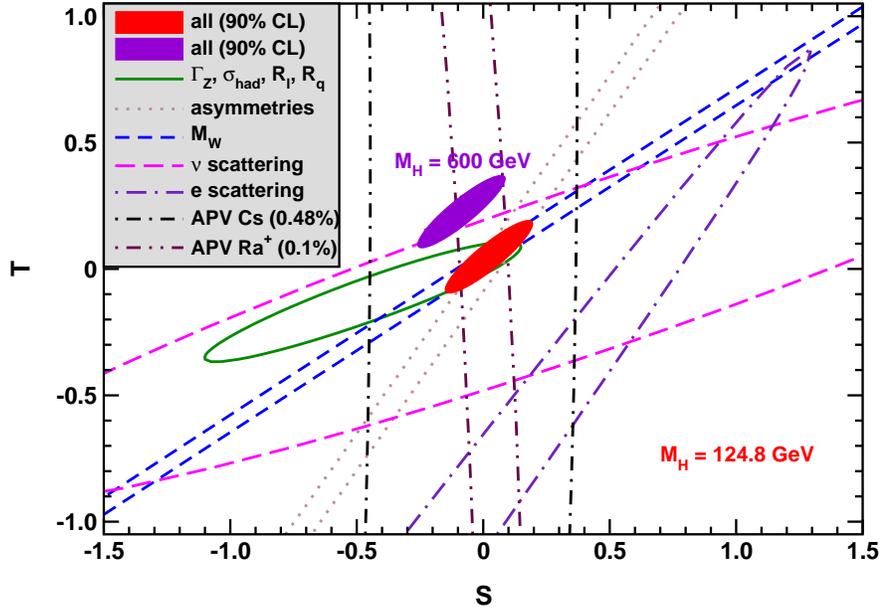}
  \caption{1~$\sigma$ constraints on $S$ and $T$ from various inputs.  
  The contours assume $M_H = 124.8$~GeV except for the upper (violet) one for all data 
  which is for $M_H = 600$~GeV.
  The contour labeled APV Ra$^+$ refers to a {\em future\/} experiment on a single trapped Ra ion
  which is in preparation at the KVI in Groningen~\cite{Jungmann:2012zz}.
  The atomic structure of Ra$^+$ is alkali-like so that the atomic theory parallels that of Cs, 
  but due to its greater neutron excess Ra constrains a linear combination of $S$ and $T$ which is
  different from Cs and quite orthogonal to the $M_W$ and $\sin^2\theta_W$ contours.
\label{ST}}
\end{figure}

\section{New Physics Interpretations}

The EW precision tests also set strong constrains on models of new physics.
{\em E.g.}, if the Higgs hints are real, an extra fermion generation 
is ruled out at the 99.6\% CL.~\cite{Kuflik:2012ai}.
This leaves us with basically three scenarios, all of which in need of some tuning and faith
(the mass spectra are generally quite similar):
(i) One ignores the collider bumps (or assigns them to something else)
and assumes $M_H \lesssim 120$~GeV (see e.g., Ref.~\cite{Dighe:2012dz});
(ii) one assumes instead $M_H \gtrsim 450$~GeV~\cite{Buchkremer:2012yy};
(iii) or one accepts $M_H \approx 125$~GeV and introduces new physics 
beyond a fourth generation, such as an extra Higgs doublet~\cite{Bellantoni:2012ag}.

\begin{figure}
  \includegraphics[height=.46\textheight]{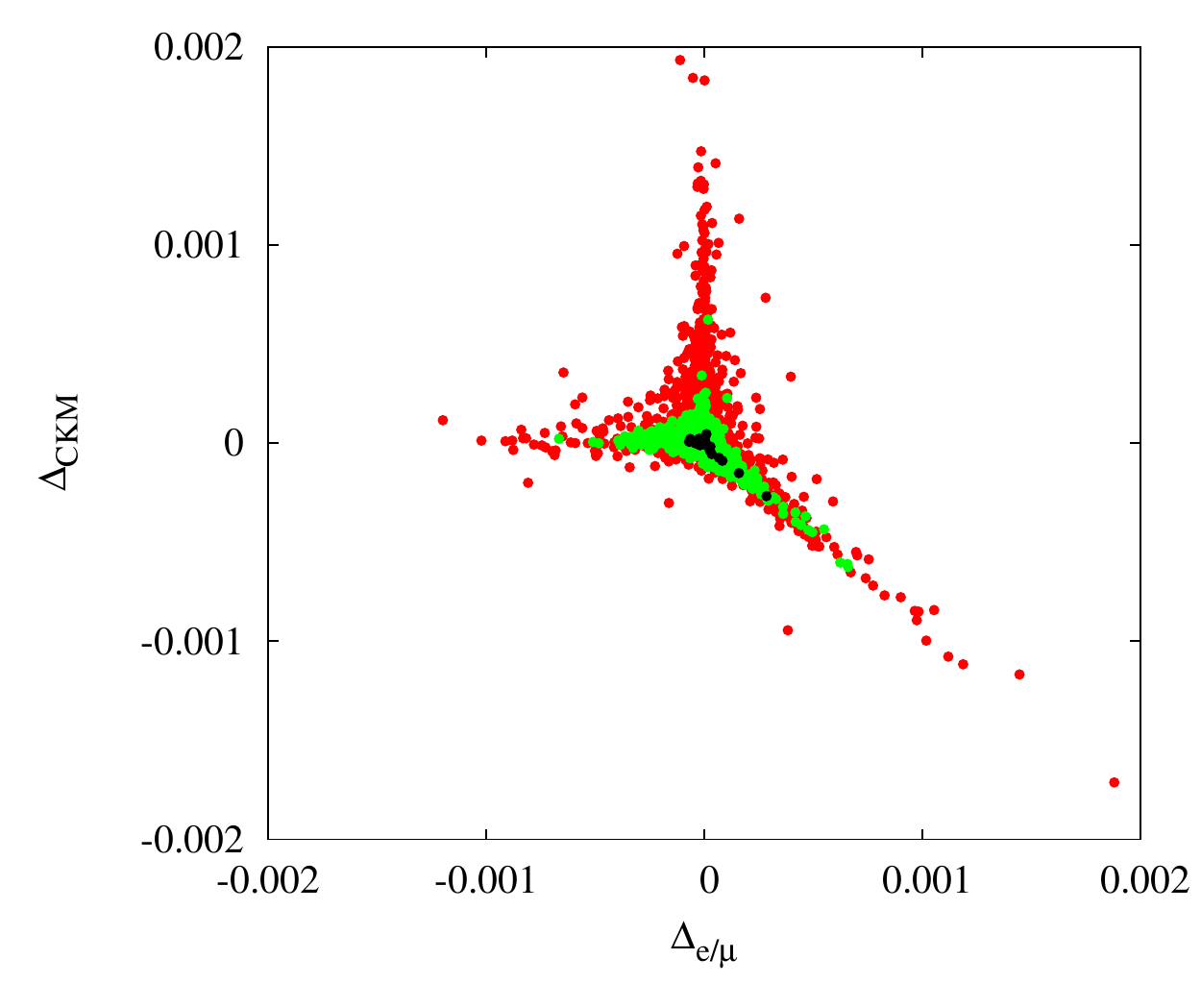}
  \caption{Scatter plot~\cite{Bauman:2012fx} of MSSM points satisfying the $\Delta_{\rm CKM}$ and 
  $\Delta_{e/\mu}$ constraints.  
  Points satisfying in addition the EW precision data including (excluding) LHC bounds are shown in
  black (green). 
  $\Delta_{\rm CKM}$ is enhanced when there is a large difference between the masses 
  of the first generation squarks and the second generation sleptons.
  Similarly, $\Delta_{e/\mu}$ is enhanced when the first and
  second generation slepton masses are significantly split.
\label{SUSYCC}}
\end{figure}

More generally, whenever the new physics is rather heavy and mostly affects 
the gauge boson self-energies, one can parametrize it in terms of the oblique 
parameters $S$ and $T$~\cite{Peskin:1991sw}
(a third parameter, $U$, is usually small).
The constraints on $S$ and $T$ from various data sets are shown in Fig.~\ref{ST}
(where $U = 0$ is assumed).

The observables discussed so far are mostly related to the weak neutral current, but
tests of charged current universality can also provide information on new physics.
Denoting any deviation from the unitarity of the first row of the CKM quark mixing matrix by 
$\Delta_{\rm CKM} \equiv |V_{ud}|^2 + |V_{us}|^2 + |V_{ub}|^2 - 1$, 
and the relative deviation from lepton universality in $\pi^+ \to \ell^+ \nu_\ell (\gamma)$ decays
($\ell = e,\mu$) by $\Delta_{e/\mu}$, one finds for the minimal supersymmetric standard 
model (MSSM) the results in Fig.~\ref{SUSYCC}. 

\section{Conclusions}

Precision tests have reached per-mille and sub per-mille accuracy in derived quantities. 
The data are in very good agreement with the SM 
with the only tantalizing deviation sitting in $a_\mu$. 
When combined with the absence of any observation challenging the SM at the LHC, 
this provides tight constraints on new physics and it becomes increasingly likely 
that its energy scale is separated from the SM by at least a little hierarchy. 

%%%%%%%%%%%%%%%%%%%%%%%%%%%%%%%%%%%%%%%%%%%%%%%%
%% BACKMATTER
%%%%%%%%%%%%%%%%%%%%%%%%%%%%%%%%%%%%%%%%%%%%%%%%

\begin{theacknowledgments}
It is a pleasure to thank the organizers of CIPANP 2012 for the invitation to an enjoyable conference.
I would also like to thank Sky Bauman, Leo Bellantoni, Jonathan Heckman, Paul Langacker, and 
Michael Ramsey-Musolf for collaboration on some of the topics presented here.
This work was supported by the CONACyT projects 82291--F and 15 1234.
\end{theacknowledgments}

\bibliographystyle{aipproc}   % if natbib is available

\bibliography{sample}

\IfFileExists{\jobname.bbl}{}
 {\typeout{}
  \typeout{******************************************}
  \typeout{** Please run "bibtex \jobname" to optain}
  \typeout{** the bibliography and then re-run LaTeX}
  \typeout{** twice to fix the references!}
  \typeout{******************************************}
  \typeout{}
 }

\end{document}